\documentclass{elsart}
\usepackage{bm}

\def\dt{{\rm d}t}
\def\d{{\rm d}}
\def\dx{{\rm d}x}
\def\norm#1{\left|\mkern-2mu\left|#1\right|\mkern-2mu\right|}

\begin{document}

\journal{Physica D}
\pubyear{2004}
\volume{}

\begin{frontmatter}

\title{Extensivity of two-dimensional turbulence}

\author{Chuong V. Tran\thanksref{label1}},
\ead{chuong@math.ualberta.ca}
\author{Theodore G. Shepherd},
\ead{tgs@atmosp.physics.utoronto.ca}
\author{Han-Ru Cho}
\ead{cho@rainbow.atm.ncu.edu.tw}

\address{Department of Physics, University of Toronto,
60 St. George Street, Toronto, ON, Canada, M5S 1A7}

\thanks[label1]{Present affiliation: Department of Mathematical and
Statistical Sciences, University of Alberta, Edmonton, Alberta, Canada,
T6G 2G1}

\begin{abstract}

This study is concerned with how the attractor dimension of the 
two-dimensional Navier--Stokes equations depends on characteristic 
length scales, including the system integral length scale, the forcing 
length scale, and the dissipation length scale. Upper bounds on the 
attractor dimension derived by Constantin--Foias--Temam are analysed. 
It is shown that the optimal attractor-dimension estimate grows 
linearly with the domain area (suggestive of extensive chaos), for 
a sufficiently large domain, if the kinematic viscosity and the 
amplitude and length scale of the forcing are held fixed. For 
sufficiently small domain area, a slightly ``super-extensive''
estimate becomes optimal. In the extensive regime, the  
attractor-dimension estimate is given by the ratio of the domain 
area to the square of the dissipation length scale defined, on 
physical grounds, in terms of the average rate of shear. This 
dissipation length scale (which is not necessarily the scale at 
which the energy or enstrophy dissipation takes place) can be 
identified with the dimension correlation length scale, the square 
of which is interpreted, according to the concept of extensive chaos, 
as the area of a subsystem with one degree of freedom. Furthermore, 
these length scales can be identified with a ``minimum length scale'' 
of the flow, which is rigorously deduced from the concept of 
determining nodes. 

\end{abstract}

\begin{keyword}

Global attractor \sep Navier--Stokes equations \sep Extensive chaos
\sep Characteristic lengths
\PACS 47.52.+j \sep 05.45.Jn \sep 47.17.+e \sep 47.27.Gs

\end{keyword}

\end{frontmatter}

\section{Introduction}
It has become a classical idea that the attractor dimension---a
measure of dynamical complexity---of a certain class of dissipative
physical systems scales linearly with the volume of the system, a
property known as ``extensive chaos"  
\cite{Cross93,Greenside95,Hohenberg89}.
This central idea of extensive chaos is physically interpreted and best
appreciated with the following heuristic argument. In a dissipative
system there exists a dissipation length scale $\ell$, below which
all modes are damped; only modes with length scales above the
dissipation scale are dynamically active. Intuitively, the attractor
dimension is approximately equal to the number of active degrees of
freedom. If other parameters can somehow be controlled so
that $\ell$ remains fixed while the volume $V$ of the system of typical
length scale $L$ is increased, then the number of active modes grows
as $(L/\ell)^d$ where $d$ is the spatial dimension of the system---i.e.
the degrees of freedom grow linearly with $V$. For demonstrative
purposes let us consider a dissipative dynamical system defined in
a (Hilbert) space $H(\Omega)$ of doubly-periodic functions on
$\Omega=(0,2\pi L)\times(0,2\pi L)$. This space is spanned by
$\{\cos L^{-1}(k_1x_1+k_2x_2),\sin L^{-1}(k_1x_1+k_2x_2)\}$
where $k_1$ and $k_2$ are integers. A basic mode of wavevector
$(k_1,k_2)^T$ above the length scale $\ell$ satisfies
\begin{eqnarray}
(k_1^2+k_2^2)^{1/2}\leq L/\ell.\nonumber
\end{eqnarray}
Modes satisfying the above inequality can be geometrically identified
as lattice sites inside the circle of radius $L/\ell$, the number of  
which is approximately the area of the circle (for a sufficiently 
large radius). It follows that the number of modes above the length 
scale $\ell$ is proportional to $(L/\ell)^2$ which is proportional 
to the domain area. Hence, extensive behaviour might be expected to 
be common among dissipative dynamical systems for which the dissipation 
length scale is independent of the system size. (Thus, the problem
of proving that a system is extensively chaotic might be restricted
to showing that its dissipation length scale is size-independent.)  
Although expected to be common, extensive behaviour has only been 
partly validated for a few cases. On the theoretical side, for example, 
extensive behaviour has been suggested for the 2D Navier-Stokes 
equations on an elongated domain for a special class of the driving 
force \cite{Ziane97} and for the one-dimensional complex Ginzburg-Landau
equation \cite{Aronson85}.\footnote{In those works, an upper bound 
on the attractor dimension that scales linearly with the domain ``volume'' 
is derived. Strictly speaking, such an upper bound does not fully 
establish extensivity, without a lower bound that behaves in the same 
manner. For the 2D Navier-Stokes equations on an elongated (square) 
doubly periodic domain driven by a simple forcing, a lower bound on
the global attractor that scales linearly with the domain area is
derived by Babin and Vishik \cite{Babin83} (Liu \cite{Liu93}).} 
On the experimental side, it has been shown
numerically that the Kuramoto--Sivashinsky model \cite{Manneville85}
and the Miller-Huse model \cite{O'Hen96} behave extensively.

Another physical interpretation of the idea of extensive chaos,  
according to its originators, is that large systems become complex 
with increasing size in a simple way by replicating weakly-interacting 
and statistically similar subsystems of some characteristic size
\cite{Cross93,Greenside95,Hohenberg89}. In other words, an extensively
chaotic system is a system such that there are no interesting collective
effects with growing size, as might be expected if small dynamical units
somehow bind themselves into a larger effective unit with fewer
total degrees of freedom. The characteristic size is identified as the
dimension correlation length
\cite{Cross93,Egolf94a,Egolf94b,Greenside95,Hohenberg89}, defined as
$\xi_\delta\equiv (V/D)^{1/d}$, where $V$ is the volume of the system,
$D$ is the (global) attractor dimension, and $d$ is the dimension of the
system's physical space. Hence, a subsystem with volume
$\xi_\delta^d=V/D$ is interpreted as a dynamical unit with one degree
of freedom. Clearly, if $\xi_\delta$ remains fixed as $V$ is varied, 
then $D\propto V$ and the system is extensive. So in this case the 
problem of proving extensivity amounts to showing that $\xi_\delta$ is
size-independent.

In this study we examine the extensivity of two-dimensional turbulence.
The two standard estimates of attractor dimension derived by 
Constantin--Foias--Temam \cite{Constantin85,Constantin88b} for the 
Navier--Stokes equations with doubly periodic boundary condition are 
analysed. It is shown that one of the estimates, when modified to take 
into account the forcing length scale, is optimal, from the perspective 
of extensive chaos. Moreover, the modification makes the physical 
significance of the forcing length scale more apparent. In particular, 
a forcing with fixed amplitude, when applied at progressively smaller 
length scales, injects a smaller amount of energy, leading to an 
increase in the dissipation length scale (decrease of Reynolds number 
and of attractor dimension). The other estimate, which is slightly 
super-extensive, turns out to be optimal when the system's length
scale is sufficiently small. We also address the physical
relevance of the dimension correlation length, and its interpretation
as the size of relatively independent subsystems in an extensively
chaotic mother system. In fact, the dimension correlation length
can be identified with the dissipation length as defined, on physical
grounds, by Hohenberg and Shraiman \cite{Hohenberg89}. Furthermore,
this length scale is equivalent to a ``minimum length scale'' rigorously
deduced from the concept of determining nodes \cite{Foias84,Friz01}.

\section{Attractor-dimension estimates for 2D turbulence}
The asymptotic behaviour of the solutions of the forced two-dimensional
Navier--Stokes equations:
\begin{eqnarray}
\label{NS}
\frac{\partial u}{\partial t}+(u\cdot\nabla)u+\nabla p &=&\nu\Delta  
u+f(x),
~~~~ \nabla\cdot u=0, ~~~~ x\in\Omega\subset\mathcal{R}^2,
\end{eqnarray}
has been the subject of intense research for the past 30 years
\cite{Babin83,Constantin88a,Constantin85,Constantin88b,Foias83,Foias79,Ladyzhenskaya69,Temam83,Tran03a,Tran02a,Ziane97}. The existence of a
bounded finite-dimensional (Hausdorff or fractal dimension) global
attractor has been established \cite{Foias79,Ladyzhenskaya69}, and
estimates on its dimension calculated
\cite{Babin83,Constantin85,Constantin88b}. The idea behind this
calculation is that if an arbitrary $m$-dimensional volume element of  
the phase space asymptotically compresses to zero, then the number of  
positive Lyapunov exponents of the system and the Hausdorff (as well 
as the fractal) dimension of the global attractor cannot exceed $m$. 
The greatest lower bound on all such possible $m$ is then an upper 
bound on the attractor dimension.\footnote{By definition, the 
Hausdorff dimension of a set is always bounded from above by its
fractal dimension and the upper bound can in cases be strict. 
Nevertheless, the greatest lower bound on $m$ derived in this 
manner is an upper bound on both the fractal and Hausdorff dimensions
of the global attractor (cf. \cite{Chepyzhov01,Hunt96}; \cite{Robinson01}, 
Appendix B). Hence, there is no need to distinguish between 
the two dimensions in this context and the term ``attractor dimension'' 
hereafter may refer to either one.} The book of Temam \cite{Temam97}
provides details of this mathematical technique and its application to
various dissipative dynamical systems. Two standard attractor-dimension 
estimates are obtained when doubly periodic solutions on
$\Omega=(0,2\pi L)\times(0,2\pi L)$ of (\ref{NS}) are considered:
\begin{eqnarray}
\label{estimate1}
D&\le&c_1G,\\
\label{estimate2}
D&\le&c_2G^{2/3}\left(1+\ln G\right)^{1/3},
\end{eqnarray}
where $D$ is the attractor dimension, $c_1$ and $c_2$ are absolute
constants, and $G$ is the generalized Grasshof number defined by
\begin{eqnarray}
\label{Grashof}
G&\equiv&\frac{\norm f}{\nu^2\lambda_1}.
\end{eqnarray}
In (\ref{Grashof}) $\norm\cdot$ is the $\mathcal{L}^2$-norm and
$\lambda_1=L^{-2}$ is the smallest eigenvalue of $-\Delta$ in
the solution phase space, being the space of non-divergent zero-average
vector-valued functions on $\Omega$. These estimates rely on the
asymptotic averages
\begin{eqnarray}
\label{Z}
\langle\norm u_{1}^2\rangle&\equiv&\limsup_{T\rightarrow\infty}
\frac{1}{T}\int_0^T\dt\int_{\Omega}u(-\Delta)u\,\dx,\\
\label{P}
\langle\norm u_{2}^2\rangle&\equiv&\limsup_{T\rightarrow\infty}
\frac{1}{T}\int_0^T\dt\int_{\Omega}|\Delta u|^2\,\dx.
\end{eqnarray}
Note that $\nu\langle\norm u_{1}^2\rangle$ and
$\nu\langle\norm u_{2}^2\rangle$ are, respectively, the average energy
dissipation (hence, the average energy injection) and the average
enstrophy\footnote{The quadratic quantity $\norm u_{1}^2/2$
(or just $\norm u_{1}^2$) is known in the classical theory of 
turbulence as the enstrophy.} dissipation (hence, the average enstrophy 
injection). For fixed average forcing amplitude $\norm f/L$ and fixed 
viscosity, $G\propto L^3$ so (\ref{estimate1}) and (\ref{estimate2}) are 
not linear in the domain area. Hence, these attractor-dimension estimates
are not suggestive of extensive chaos although (\ref{estimate2}) is 
``essentially'' extensive, only slightly ``super-extensive'' by a 
logarithmic term. 

However an intermediate step towards (\ref{estimate1}) is
\begin{eqnarray}
\label{estimate11}
D&\le&c_1\frac{\langle\norm u_{1}^2\rangle^{1/2}}{\nu\lambda_1^{1/2}}.
\end{eqnarray}
The energy equation gives the straightforward asymptotic inequality
$\langle\norm u_{1}^2\rangle^{1/2}\le\norm f/(\nu\lambda_1^{1/2})$,
from which (\ref{estimate11}) then implies (\ref{estimate1}).
As noted independently by Robinson \cite{Robinson03}
and Tran \cite{Tran01}, the two attractor-dimension estimates  
(\ref{estimate1}) and (\ref{estimate2}) do not take into account the 
spatial scale of the forcing. A simple modification to the estimate 
of $\langle\norm u_{1}^2\rangle$ improves (\ref{estimate1}) and gives it
an explicit dependence on the forcing length scale
(see \cite{Babin83,Robinson03}; \cite{Tran01}, p.40). To this end let 
us consider the energy evolution equation
\begin{eqnarray}
\frac{1}{2}\frac{\d}{\dt}\norm u^2&=&-\nu\norm  
u_1^2+\int_{\Omega}u\,f\,\dx
=-\nu\norm u_1^2+\int_{\Omega}(-\Delta)^{1/2}u(-\Delta)^{-1/2}f\,\dx
\nonumber\\
&\le&-\nu\norm u_1^2+\norm u_1\norm f_{-1}
\le-\frac{1}{2}\nu\norm u_1^2+\frac{1}{2\nu}\norm f_{-1}^2,
\end{eqnarray}
where the two inequalities are obtained by applying the Schwarz and
Young inequalities, respectively. It follows that asymptotically
\begin{eqnarray}
\label{Zbound}
\langle\norm u_{1}^2\rangle&\le&\frac{1}{\nu^2}\norm f_{-1}^2.
\end{eqnarray}
In terms of this estimate (\ref{estimate11}) becomes
\begin{eqnarray}
\label{estimate111}
D&\le&c_1\frac{\norm f_{-1}}{\nu^2\lambda_1^{1/2}}.
\end{eqnarray}
This estimate, together with (\ref{estimate2}), which is written
explicitly in terms of the physical parameters as
\begin{eqnarray}
\label{estimate222}
D&\le&c_2\left(\frac{\norm f}{\nu^2\lambda_1}\right)^{2/3}
\left(1+\ln\frac{\norm f}{\nu^2\lambda_1}\right)^{1/3},
\end{eqnarray}
are the focus of our present analysis.

The norm $\norm f_{-1}$ implicitly contains information
about the spatial scale of $f$. For example, a monoscale forcing
$f$, i.e. $-\Delta f=\lambda_sf$ where $\lambda_s$ is an eigenvalue of
$-\Delta$, yields $\norm f_{-1}=\lambda_s^{-1/2}\norm f$. For a general
forcing one may define the forcing length scale
$\ell_s\equiv\norm f_{-1}/\norm f$ \cite{Tran01}. Note that $\ell_s$ 
is always\footnote{The case $\ell_s=L$ is pathological as the
forcing then belongs to the eigenspace of $-\Delta$ corresponding to
$\lambda_1$. This type of forcing is well known to lead to a trivial
attractor consisting of a single stable stationary solution. This fact
was first proven by Iudovich \cite{Iudovich65} and later investigated
by Marchioro \cite{Marchioro86}. A simple proof is given by
Constantin--Foias--Temam \cite{Constantin88b}.} less than $L$, so that
$\norm f_{-1}<\norm f/\lambda_1^{1/2}$. Hence, (\ref{estimate111})
is an improved estimate as compared to (\ref{estimate1}) (see 
\cite{Tran01}, p.44). This fact is independently noted by Robinson 
\cite{Robinson03}.

We note in passing that while the energy injection decreases when a
forcing with fixed $\norm f$ is applied at progressively smaller
length scales, the enstrophy injection $\int_{\Omega}(-\Delta)u\,f\,\dx$ 
does not necessarily do so. Therefore, it does not seem possible to 
apply the above trick to improve (\ref{estimate2}), which is based 
on an estimate of the enstrophy injection.

{\bf Remark 1.} The appearance of the forcing length scale in
(\ref{estimate111}) is natural. The forcing scale has been shown to play
a central role in the spectral distribution of energy and enstrophy and
of their dissipation \cite{Tran02a}. It also has non-trivial effects on
optimal conditions for nonlinear stability \cite{Tran02b}.

{\bf Remark 2.} It is a highly non-trivial problem to determine the
energy injection rate $\int_{\Omega}u\,f\,\dx$. This depends on how the
fluid responds to a particular forcing $f$ and cannot be identified
with $\norm f^2$. The estimate of $\int_{\Omega}u\,f\,\dx$ using the
Schwarz and Young inequalities, leading to (\ref{Zbound}) and eventually
to (\ref{estimate111}), provides what is essentially an upper bound on
the energy injection rate. When a forcing $f$ with fixed $\norm f$ 
is applied at progressively smaller scales, this upper bound decreases, 
leading to a decrease of the attractor-dimension estimate 
(\ref{estimate111}). The interpretation is that the decrease of the 
attractor dimension is due to the decrease of the energy injection,
leading to an increase in the dissipation length scale (decrease of
Reynolds number).

{\bf Remark 3.} It is concluded in \cite{Robinson03} that when a fixed
amount of energy is injected at progressively smaller scales then the
dimension of the attractor decreases. This conclusion seems to originate
from the (mis)identification of $\norm f^2$ with the energy injection
rate. It is correct to say that a forcing with a fixed $\norm f$ when 
applied at progressively smaller scales leads to a decrease of the 
attractor dimension, as seen above. However, such a forcing, when 
applied at progressively smaller scales, does not inject the same 
amount of energy and the interpretation in the preceding remark is 
more plausible.

{\bf Remark 4.} To clarify the arguments in the previous two remarks 
let us consider the case of monoscale forcing $f$. This forcing
gives rise to the (primary) stationary solution  
$\bar{u}=(-\nu\Delta)^{-1}f=f/\nu\lambda_s$ (see Remark 5), which is 
stable for sufficiently small $\norm{\bar u}$ \cite{Tran02b}. In the 
stable regime, the energy injection rate is
$\int_{\Omega}\bar{u}f\,\dx=\norm f^2/(\nu\lambda_s)=\norm f^2_{-1}/\nu$.
This energy injection rate indeed decreases with increasing $\lambda_s$
(decreasing forcing length scale), for a fixed $\norm f$.

{\bf Remark 5.} The fact that a monoscale forcing gives rise to a 
monoscale stationary solution is more readily deduced from the 
vorticity equation---$\partial_t\Delta\psi+\partial_x\psi\,\partial_y
\Delta\psi-\partial_y\psi\,\partial_x\Delta\psi=\nu\Delta^2\psi+g$,
where $g=\widehat{z}\cdot\nabla\times f$ ($\widehat{z}$ is the normal
to the fluid domain)---than from (\ref{NS}). If $g$ satisfies 
$\Delta g=-\lambda_sg$, then $\bar\psi=-g/(\nu\lambda_s^2)$ is a 
stationary solution because the nonlinear terms identically vanish 
for $\psi=\bar\psi$. In the traditional NS system, $(u\cdot\nabla)u$ 
is non-zero (and not even non-divergent) for a non-divergent monoscale 
$u$. However, the divergent component of $(u\cdot\nabla)u$ is balanced 
by the pressure gradient term, and its non-divergent component vanishes
for a non-divergent monoscale $u$. The latter component is described 
in the literature as the projection of the nonlinear term onto the 
non-divergent zero-average phase space, and is often denoted by $B(u,u)$. 
This projection is essentially equivalent to taking the curl of 
$(u\cdot\nabla)u$ to obtain the vorticity equation. Detailed 
treatments of $B(u,u)$ are given in \cite{Liu93,Tran02b}, where it is 
noted that the nonlinear interaction coefficient between two modes $k$ 
and $l$ contains the factor $\widehat{z}\cdot k\times l\,(|k|^2-|l|^2)$ 
[$\widehat{z}$ is the normal to the wavevector plane]. (Pedlosky 
\cite{Pedlosky87} notes that the nonlinear term in the vorticity equation
vanishes for waves with the same wavelength or with parallel wavevectors.)
Therefore, if the forcing $f$ in (\ref{NS}) consists of basic modes, for 
which the wavevectors are parallel or have the same magnitude, then 
$(-\nu\Delta)^{-1}f$ is a stationary solution. Thus, $\bar u$ is a 
stationary solution.

\section{Characteristic length scales}

There are various length scales that characterize the dynamics of a
spatially extended system. Hohenberg and Shraiman \cite{Hohenberg89}
distinguish three length scales which are associated with excitation, 
dissipation, and correlation, and suggest that it is the ratios of these
length scales to one another and to the typical system length scale  
that will determine the dynamics of the system. The excitation length 
scale is straightforward; it
is the scale on which energy is injected into the system by 
external forcing. The dissipation length scale $\ell$ characterizes 
the length scale below which all modes are damped in a finite time. 
For fluid flow where molecular viscosity alone is responsible for 
dissipation, $\ell\sim (\nu\tau)^{1/2}$ where $\tau^{-1}$ is the 
local rate of shear \cite{Hohenberg89}. This definition is applicable 
to both two- and three-dimensional flows. However, it is not the only 
definition of $\ell$ in each case. In two-dimensional flow, for 
example, Batchelor \cite{Batchelor69} and Kraichnan \cite{Kraichnan67} 
define the dissipation length scale as $\ell=(\nu^3/\eta)^{1/6}$, 
where $\eta$ is the average enstrophy flux to small scales which 
asymptotically must equal the average rate of enstrophy 
dissipation $\nu\langle\norm u^2_2\rangle/L^2$. It should be noted that 
these definitions have their roots entirely in physical and dimensional
considerations. In particular, the latter definition is based on the 
hypothesis that the enstrophy cascades to small scales ($<\ell$)
via a $k^{-3}$ energy spectrum, so that $\ell$ is approximately 
identified with the length scale corresponding to the high-wavenumber 
end of the enstrophy-cascading range \cite{Kraichnan67}. In the absence
of a direct enstrophy cascade on the global attractor, as is argued
in \cite{Tran02a}, this length scale loses its supposed physical 
relevance. The former definition, based on the local rate of shear, is
associated with the energy dissipation instead of with the enstrophy 
dissipation, thereby not suffering from this difficulty. Nevertheless, 
given that there exists no dissipation length scale (for both energy 
and enstrophy) well separated from the forcing scale for a broad 
class of forcing \cite{Tran02a}, including the one considered by the 
classical theory of turbulence \cite{Kraichnan67,Lesieur97,Pouquet75}, 
the terminology ``dissipation length scale'' may not refer to where 
in wavenumber space the dissipation of energy (or enstrophy) occurs. 
Rather, it refers, in a qualitative sense, to the length scale below 
which viscous effects dominate nonlinearity. We will see later in this 
section that this length scale is better quantified by the concept 
of determining nodes \cite{Foias84,Friz01}. In fact, it can be 
identified with a ``minimum length scale'' of the flow that can be 
rigorously deduced. 

The concept of correlation length 
has a relatively long history in fluid dynamics (dating back to early 
in the 1900s), and in dynamical systems theory in general. There has 
been more than one definition of this length scale; two that are most 
relevant to the present discussion will be briefly reviewed. The first 
and simpler one is the two-point correlation length defined in terms 
of the correlation function
\begin{eqnarray}
C(x,x')\equiv\langle(\phi(x,t)-\langle\phi\rangle_x)(\phi(x',t)
-\langle\phi\rangle_x)\rangle_t,\nonumber
\end{eqnarray}
where the angle brackets with subscript $x$ denote the spatial average,
the angle brackets with subscript $t$ denote the time average, and
$\phi(x,t)$ is some local variable. In cases where
$C(x,x')\sim\exp\{-|x-x'|/\xi\}$
as $|x-x'|\rightarrow\infty$ one can then define the two-point
correlation length $\xi$. This length scale, if it exists, may be
identified with the size of weakly-interacting subsystems. The second
correlation length is called the dimension correlation length, the
definition of which is due to Cross and Hohenberg \cite{Cross93}.
According to the idea of extensive chaos, there exists a dimension 
density $\delta$ being the ratio of the attractor fractal dimension 
(or equivalently Hausdorff dimension) $D$ to the system volume $V$,
\begin{eqnarray}
\delta\equiv\frac{D}{V}.\nonumber
\end{eqnarray}
Since $\delta$ has the physical units of inverse volume, Cross and
Hohenberg \cite{Cross93} suggest defining a dimension correlation
length $\xi_{\delta}$:
\begin{eqnarray}
\label{dimlength}
\xi_{\delta}\equiv\delta^{-1/d},
\end{eqnarray}
where $d$ is the dimension of the system's physical space. It is highly
likely that the dimension correlation length so defined might not be
a new length scale. The justification for this claim lies in the  
heuristic argument in favor of extensive chaos provided in Section 1. 
It is argued there that the attractor dimension $D$ is of the order 
of $(L/\ell)^d$, so that $\delta=\ell^{-d}$. Substituting this into 
(\ref{dimlength}) one obtains $\xi_{\delta}=\ell$. Thus, def. 
(\ref{dimlength}) might just be the familiar dissipation length scale.

The present problem basically has four physical parameters, which
can be independently controlled: the system linear length scale $L$, 
the forcing length scale $\ell_s$, the forcing
root mean square amplitude (forcing density) $|f|\equiv\norm f/L$,
and the molecular viscosity $\nu$. The first two parameters make 
obvious and natural length scales. The last two parameters can be 
grouped, on dimensional grounds, to form a third length scale. Since 
the quantity $(\nu^2/|f|)^{1/3}$ has the physical units of length, 
it is tempting to designate this quantity as a dissipation length 
scale. It turns out, however, that by grouping $\nu$, $|f|$, and 
$\ell_s$ one can form a more physically plausible length scale
which will be identified with the dissipation length scale $\ell$:
\begin{eqnarray}
\ell&=&\left(\frac{\nu^2}{|f|\ell_s}\right)^{1/2}.
\end{eqnarray}
This agrees with the definition of $\ell$ in \cite{Hohenberg89} because,
according to (\ref{Zbound}), $|f|\ell_s/\nu$ is an estimate of (strictly,  
an upper bound for) the average rate of shear 
$\langle\norm u_1^2\rangle^{1/2}/L$, thereby related to the
energy dissipation (not enstrophy dissipation). Thus, this definition 
is expected to differ from the one due to Kraichnan and Batchelor. 
To see the extent of this difference, we write out the 
explicit expressions for these length scales, denoting the one due to 
Kraichnan and Batchelor by $\ell_{KB}$ and preserving the notation
$\ell$ for the other,
\begin{eqnarray}
\ell_{KB}&\approx&\left(\frac{\nu L}{\langle\norm u_2^2\rangle^{1/2}}
\right)^{1/3},\\
\ell&\approx&\left(\frac{\nu L}{\langle\norm u_1^2\rangle^{1/2}}
\right)^{1/2}.
\end{eqnarray}
It follows that 
\begin{eqnarray}
\frac{\ell_{KB}^3}{\ell^2}&\approx&
\frac{\langle\norm u_1^2\rangle^{1/2}}{\langle\norm u_2^2\rangle^{1/2}}.
\end{eqnarray}
The right-hand side of (16) is the enstrophy dissipation length scale, which 
according to the classical theory \cite{Batchelor69,Kraichnan67} would
presumably be $\ell_{KB}$. This would imply $\ell_{KB}\approx\ell$.
However, in the usual case of a spectrally localized forcing around 
a forcing length scale $\ell_s$, such that the energy and enstrophy 
injection rates are related by $\langle\int_\Omega u\,f\,\dx\rangle=
\ell_s^2\langle\int_\Omega(-\Delta)u\,f\,\dx\rangle$, one has 
$\langle\norm u_1^2\rangle^{1/2}/\langle\norm u_2^2\rangle^{1/2}=\ell_s$
\cite{Tran02a}. This result implies that $\ell_{KB}\approx
(\ell^2\ell_s)^{1/3}$, which makes the difference between $\ell_{KB}$ 
and $\ell$ significant if $\ell\ll\ell_s$.

Closely related to the notion of dissipation length scale are the 
concepts of determining modes, nodes, and finite-volume elements
\cite{Foias83,Foias84,Friz01,Jones92,Jones93}, from which a 
``minimum length scale'' of the flow can be rigorously deduced. 
Here we restrict our discussion to the concept of determining nodes,
which is introduced by Foias and Temam \cite{Foias84} and for which 
a sharp (sharpest to date) estimate of such a length scale is derived
by Friz and Robinson \cite{Friz01}. A simplified version of Friz and 
Robinson's result can be stated as follows. Let $\{x_i\}_{i=1}^n$ be 
a set of $n$ lattice sites (nodes) equally spaced in the domain 
$\Omega$, so that the lattice spacing is $L/n^{1/2}$. Suppose that 
$u$ and $v$ are two trajectories on the global attractor and that 
$u$ and $v$ agree at the lattice sites, i.e. $u(x_i)=v(x_i)$, for 
$i=1,2,\cdots,n$. If this condition is satisfied, where $n$ is of 
the order of the attractor dimension $D$ and the forcing is analytic,
then $u=v$. This result means that a trajectory on the global 
attractor is completely determined by its values at $D$ discrete 
nodes, hence the term ``determining nodes''. Obviously, the lattice 
spacing $L/D^{1/2}$ coincides with the dimension correlation length 
scale $\xi_\delta$ previously mentioned. 
   
In terms of the length scales described in the preceding paragraphs,
the results of the last section can be written as
\begin{eqnarray}
\label{nsdim3}
D&\le&\left(\frac{L}{\ell}\right)^2,\\
\label{nsdim4}
D&\le&\left(\frac{L^3}{\ell^2\ell_s}\right)^{2/3}
\left(1+\ln\frac{L^3}{\ell^2\ell_s}\right)^{1/3},
\end{eqnarray}
where a constant of order unity has been dropped from each expression.
A couple of remarks are in order. First, not only does expression
(\ref{nsdim3}) agree with the idea of extensive chaos, but it also takes
the form suggested by the heuristic argument in the introductory  
section. Second, it is clear that the logarithmic correction term in
expression (\ref{nsdim4}) surely diverges as $L\rightarrow\infty$;
however, (\ref{nsdim4}) may well be optimal for small domain size. In
that case the system is slightly ``super-extensive". When $L$ is
sufficiently large, the bound on $D$ behaves extensively as  
(\ref{nsdim3}) becomes optimal. In this regime, the attractor-dimension
estimate grows linearly with the domain area, provided $\ell$ is fixed.
Constancy of $\ell$ in a varying domain size requires fixed $|f|$,
$\nu$, and $\ell_s$. A fixed $\ell_s$ means that the non-dimensional 
characteristic wavenumber of the forcing increases in exact proportion
with $L$. Note that for the extensive regime (\ref{nsdim3}) implies
that $\ell\le L/D^{1/2}$, so $\ell$ can be identified with the 
``minimum length scale'' in the sense of Foias and Temam and 
of Friz and Robinson. For the super-extensive regime $\ell_{KB}$ 
plays the role of this length scale if the logarithmic correction is 
ignored. The correspondence between $\ell_{KB}$ and the rigorous 
``minimum length scale'' (ignoring the logarithmic correction) has 
previously been noted by Friz and Robinson \cite{Friz01}. The length 
scale that separates the two regimes may be called the extensivity 
length scale \cite{Tran01}. Letting $L_e$ denote this length, a 
comparison of (\ref{nsdim3}) and (\ref{nsdim4}) gives
\begin{eqnarray}
\left(\frac{L_e}{\ell}\right)^2&\sim&\left(\frac{L_e^3}{\ell^2\ell_s}
\right)^{2/3}\left(1+\ln\frac{L_e^3}{\ell^2\ell_s}\right)^{1/3}.
\nonumber
\end{eqnarray}
It follows that
\begin{eqnarray}
L_e&\sim&\exp\left\{\frac{\ell_s^2}{3\ell^2}\right\}(\ell^2\ell_s)^{1/3},
\end{eqnarray}
provided $\ell\ll \ell_s$. It can be confirmed that $L_e>\ell_s$, so a
non-trivial extensivity length scale does exist.

Finally, the dimension correlation length as defined by Cross and
Hohenberg \cite{Cross93} and advanced by Egolf \cite{Egolf94a,Egolf94b}
is just the newly defined dissipation length $\ell$, which has the
physical significance as discussed above. In particular, $\ell$ 
represents the spacing of determining nodes, and in the absence of 
a dissipation range well separated from the forcing region \cite{Tran02a}, 
the terminology associated with $\ell$ finds itself a better 
justification in this new sense.

\section{Conclusion}

We have analysed the two standard attractor-dimension estimates
of the two-dimensional Navier--Stokes equations, derived by
Constantin--Foias--Temam. It is shown that one of the estimates
grows linearly with the domain area, for sufficiently large systems,
when the kinematic viscosity and forcing density and its length scale  
are held fixed. This is consistent with the central idea of extensive
chaos: namely the linear scaling of complexity, as characterized by the
attractor dimension, with domain area. This qualifies (\ref{estimate111})
as the optimal estimate for large systems. For small systems,
for which the domain length scale is below a threshold called
the extensivity length scale, the estimate (\ref{estimate222}) becomes  
optimal. In this regime, the attractor dimension is slightly 
super-extensive, essentially by a logarithmic factor.

In the extensive regime, the upper bound on the attractor dimension is  
given by the ratio of the domain area to the square of the dissipation 
length scale, which is defined, on physical grounds, in terms of the 
average shear. This implies an equivalence between the dissipation 
length scale and the dimension correlation length scale. The latter
has been suggested to be a new characteristic length scale for 
extensively chaotic systems, but it seems likely that it is just 
another way of determining the dissipation length scale.

{\bf Acknowledgements}

The work reported here represents part of CVT's Ph.D. thesis at the
University of Toronto, which was supported by University of Toronto
Open Fellowships and Department of Physics Burton Fellowships. TGS
would like to acknowledge support from the Natural Sciences and
Engineering Research Council and the Meteorological Service of Canada.
Constructive comments from a reviewer were very much appreciated.

%\bibliography{refs}

\begin{thebibliography}{10}

\bibitem{Aronson85} I.S. Aranson, A.V. Gaponov-Grekhov, M.I. Rabinovich,
The development of chaos in dynamic structure ensembles, Sov. Phys. JETP
62 (1985) 52--59.

\bibitem{Babin83} A.V. Babin, M.I. Vishik, Attractors of partial
differential equations and estimate of their dimensions, Russian
Math. Surveys 38:4 (1983) 151--213.

\bibitem{Batchelor69} G.K. Batchelor, Computation of the energy
spectrum in homogeneous two-dimensional turbulence, Phys. Fluids 12
(II) (1969) 233--239.

\bibitem{Chepyzhov01} V.V. Chepyzhov, A.A. Ilyin, A note on the 
fractal dimension of attractors of dissipative dynamical systems,
Nonlinear Anal. 44 (2001) Ser. A: Theory Methods, 811--819.

\bibitem{Constantin88a} P. Constantin, C. Foias, Navier--Stokes
Equations, University of Chicago Press, Chicago, 1988.

\bibitem{Constantin85} P. Constantin, C. Foias, R. Temam, Attractors
representing turbulent flows, Mem. Am. Math. Soc. 53 (314) (1985)
1--67.

\bibitem{Constantin88b} P. Constantin, C. Foias, R. Temam, On the
dimension of the attractors in two-dimensional turbulence, Physica D
30 (1988) 284--296.

\bibitem{Cross93} M.C. Cross, P.C. Hohenberg, Pattern formation outside
of equilibrium, Rev. Mod. Phys. 65 (1993) 851--1112.

\bibitem{Egolf94a} D.A. Egolf, Characterization of extensively chaotic
states and transitions, Ph.D. thesis, Department of Physics,
Duke University, Durham, NC, 1994.

\bibitem{Egolf94b} D.A. Egolf and H.S. Greenside, Relation between
fractal dimension and spatial correlation length for extensive chaos,
Nature 369 (1994) 129--131.

\bibitem{Foias83} C. Foias, O.P. Manley, R. Temam, Y.M. Treve,
Asymptotic analysis of the Navier--Stokes equations, Physica D 9 (1983)
157--188.

\bibitem{Foias79} C. Foias, R. Temam, Some analytic and geometric
properties of the solutions of the Navier--Stokes equations,
J. Math. Pure Appl. 58 (1979) 339--369.

\bibitem{Foias84} C. Foias, R. Temam, Determination of the solutions 
of the Navier--Stokes equations by a set of nodal values, Math. Comp.
43 (1984) 117--133.

\bibitem{Friz01} P.K. Friz, J.C. Robinson, Parametrising the attractor
of the two-dimensional Navier--Stokes equations with a finite set of
nodal values, Physica D 148 (2001) 201--220.

\bibitem{Greenside95} H.S. Greenside, Spatiotemporal chaos in large
systems: the scaling of complexity with size. Preprints of the Montreal
workshop of the CRM (Centre de Recherche en Mathematiques) under the
title ``Semi-analytic methods for the Navier-Stokes equations", Oct.  
1995.

\bibitem{Hohenberg89} P.C. Hohenberg, B.I. Shraiman, Chaotic behaviour
of an extended system, Physica D 37 (1989) 109--115.

\bibitem{Iudovich65} V.I. Iudovich, Example of the generation of a
secondary stationary or periodic flow when there is loss of stability
of the laminar flow of a viscous incompressible fluid, J. Appl. Math.
Mech. 29 (1965) 527--544.

\bibitem{Jones92} D.A. Jones, E.S. Titi, Determining finite volume 
elements for the 2D Navier--Stokes equations, Physica D 60 (1992) 165--174.

\bibitem{Jones93} D.A. Jones, E.S. Titi, Upper bounds on the number 
of determining modes, nodes, and volume elements for the Navier--Stokes
equations, Indiana Univ. Math. J. 42 (1993) 875--887.

\bibitem{Kraichnan67} R.H. Kraichnan, Inertial ranges in two-dimensional
turbulence, Phys. Fluids 10 (1967) 1417--1423.

\bibitem{Ladyzhenskaya69} O.A. Ladyzhenskaya, The Mathematical Theory of
Viscous Incompressible Flow, Gordon and Breach, New York, 2nd ed., 1969.

\bibitem{Lesieur97} M. Lesieur, Turbulence in Fluids, 3rd edition, 
Kluwer, Dordrecht, 1997.

\bibitem{Liu93} V.X.S. Liu, A sharp lower bound for the Hausdorff
dimension of the global attractors of the 2D Navier--Stokes equations,
Comm. Math. Phys. 158 (1993) 327--339.

\bibitem{Manneville85} P. Manneville, Lyapunov exponents for the
Kuramoto-Sivashinsky model, Macroscopic modeling of turbulent flows
(O. Pironneau, ed.), Lecture Notes in Physics, vol. 230,
Springer--Verlag, New York, 1985, 319--326.

\bibitem{Marchioro86} C. Marchioro, An example of absence of turbulence
for any Reynolds number, Comm. Math. Phys. 105 (1986) 99--106.

\bibitem{O'Hen96} C. O'Hern, D. Egolf, H. Greenside, Lyapunov spectral
analysis of a nonequilibrium Ising-like transition, Phys. Rev. E 53
(1996) 3374--3386.

\bibitem{Hunt96} B.R. Hunt, Maximum local Lyapunov dimension bounds the box
dimension of chaotic attractors, Nonlinearity 9 (1996) 845--852.

\bibitem{Pedlosky87} J. Pedlosky, Geophysical Fluid Dynamics, 2nd edition, 
Springer--Verlag, New York, 1987.

\bibitem{Pouquet75} A. Pouquet, M. Lesieur, J.C. Andr\'e, and C. Basdevant,
Evolution of high Reynolds number two-dimensional turbulence, J. Fluid
Mech. 72 (1975) 305--319.

\bibitem{Robinson03} J.C. Robinson, Low dimensional attractors arise
from forcing at small scales, Physica D 181 (2003) 39--44.

\bibitem{Robinson01} J.C. Robinson, Infinite-Dimensional Dynamical
Systems, Cambridge University Press, New York, 2001.

\bibitem{Temam97} R. Temam, Infinite-Dimensional Dynamical Systems in
Mechanics and Physics, Springer--Verlag, New York, 2nd ed., 1997.

\bibitem{Temam83} R. Temam, Navier--Stokes Equations and Nonlinear  
Functional Analysis, SIAM, Philadelphia, 1983.

\bibitem{Tran01} C.V. Tran, Extensive Chaos and Complexity of
Two-Dimensional Turbulence, Ph.D. Thesis, University of Toronto, 2001.

\bibitem{Tran03a} C.V. Tran, J.C. Bowman, On the dual cascade in 
two-dimensional turbulence, Physica D 176 (2003) 242--255.

\bibitem{Tran02a} C.V. Tran, T.G. Shepherd, Constraints on the spectral
distribution of energy and enstrophy dissipation in forced  
two-dimensional turbulence, Physica D 165 (2002) 199--212.

\bibitem{Tran02b} C.V. Tran, T.G. Shepherd, H.-R. Cho, Stability of
stationary solutions of the forced Navier--Stokes equations on the
two-torus, Discrete Contin. Dyn. Syst. Ser. B 2 (2002) 483--494.

\bibitem{Ziane97} M. Ziane, Optimal bounds on the dimension of the
attractor of the Navier--Stokes equations, Physica D 105 (1997) 1--19.

\end{thebibliography}

\end{document}